\documentstyle[seceq]{ptptex}

\def\be{\begin{equation}}
\def\ee{\end{equation}}
\def\tr{{\rm tr}}
\def\l{\left}
\def\r{\right}
\def\ba{\begin{array}}
\def\ea{\end{array}}
\def\bea{\begin{eqnarray}}
\def\eea{\end{eqnarray}}
\def\nn{\nonumber}

\def\p{\partial}
\def\f{\frac}
\def\tr{{\rm tr}}
\def\bra{\langle}
\def\ket{\rangle}
\def\vx{\vspace{0.3cm}}

\notypesetlogo  

\title{Left-Right Symmetric Model from Geometric Formulation of
Gauge Theory in $M_4 \times Z_2 \times Z_2$}

\author{
Gaku Konisi,\footnote{E-mail: konisi@ka2.so-net.ne.jp}
Ziro Maki$^{*,}$,\footnote{E-mail: ziromaki@phys.kindai.ac.jp}
Mikio Nakahara$^{*,}$\footnote{E-mail: nakahara@math.kindai.ac.jp} and
Takesi Saito\footnote{E-mail: tsaito@jpnyitp.yukawa.kyoto-u.ac.jp}}
\inst{
Department of Physics, Kwansei Gakuin University\\
Nishinomiya 662-8501, Japan\\
$^{*}$Department of Physics, Kinki University, Higashi-Osaka 577-8502, Japan
}

\recdate{
}

\abst{%
The left-right symmetric model (LRSM) with gauge group
$SU(2)_{L} \times SU(2)_{R}
\times U(1)_{B-L}$ is reconstructed from the geometric formulation
of gauge theory in $M_4 \times Z_2 \times Z_2$ where $M_4$ is
the four-dimensional Minkowski space and $Z_2 \times Z_2$ the discrete
space with four points. The geometrical structure of this model becomes
clearer compared with other works based on noncommutative geometry.
As a result, the Yukawa coupling terms and the Higgs potential are
derived in more restricted forms than in the standard LRSM.
}

\begin{document}

\maketitle

\section{Introduction}

Recently gauge theories on the space $M_4 \times Z_N$, where $M_4$ is
the four-dimensional Minkowski space and $Z_N$ the discrete space with
$N$ points, have been formulated in terms of noncommutative geometry
(NCG) by Connes \cite{ref:1} and in various alternative versions of this
theory\cite{ref:2} to apply to spontaneouly broken gauge theories based on the
Higgs mechanism. In these theories the Higgs fields are regarded as
gauge fields along the discrete space $Z_N$.

All existing approaches based on NCG, however, appear to be too
algebraic, rather than geometric, and hence their geometrical meanings
have not been clarified in depth. Recently an alternative formulation of the
gauge theory in $M_4 \times Z_N$ has been proposed from purely geometrical
point of view, without employing the entire context of NCG.\cite{ref:3}
In this approach the Higgs fields are introduced as mapping functions
between any pair of vector fields, each of which is defined independently
on a sheet in the $N$-sheeted space-time. This theory
has been applied to several physical models and their geometrical
structures have been clarified.\cite{ref:4}

In the present paper, we reconstruct
the left-right symmetric model $SU(2)_{L} \times SU(2)_{R}
\times U(1)_{B-L}$ (LRSM)\cite{ref:5} in terms of
the geometric formulation of the gauge theory in $M_4 \times
Z_2 \times Z_2$. Our approach leads to the LRSM quite successfully
and the geometrical structure of the model becomes clearer
compared to other works based on NCG.\cite{ref:6} 

It has long been suspected that the standard model is not a final theory.
Rather it will be replaced by some more unified theory at higher energies.
The recent discovery of neutrino
flavor oscillation\cite{ref:7} stimulated our imagination to find
such a theory. The LRSM is one of such
unified theories that generates the light neutrino mass by the seesaw
mechanism.\cite{ref:8}
Let us summarize the LRSM in its simplest form in the following.
The quark sectors are not considered here and only the lepton sector
of the first generation will be taken into account.
The standard Lagrangian of the LRSM is given by\cite{ref:9}
\be
{\cal L} = {\cal L}_F + {\cal L}_Y + {\cal L}_B,
\ee
where
\bea
{\cal L}_F
&=&i \bar{l}_L \gamma^{\mu}\l(\p_{\mu}+i\f{1}{2} g_1 B_{\mu}
-i g_2 W^L_{\mu} \r) l_L \nn\\
& &+ i \bar{l}_R \gamma^{\mu}\l(\p_{\mu}+i\f{1}{2} g_1 B_{\mu}
-i g_2 W^R_{\mu} \r) l_R\\
{\cal L}_Y&=& -\bar{l}_L(f \phi+\tilde{f} \tilde{\phi}) l_R +{\rm h.c.}\nn\\
& & -i l^T_L C \tau_2 h \Delta_L l_L  +{\rm h.c.}\nn\\
& & -i l^T_R C \tau_2 h \Delta_R l_R  +{\rm h.c.},\\
{\cal L}_B &=& \tr\ |D_{\mu} \Delta_L|^2 + \tr\ |D_{\mu}\Delta_R|^2
+ \tr |D_{\mu} \phi|^2\nn\\
& &+ \mbox{Yang-Mills terms of }B_{\mu}, W_{\mu}^L, W_{\mu}^R \nn\\
& &-V(\mbox{Higgs potential of }\phi, \Delta_L, \Delta_R),
\eea
and
\bea
l_L &=& \l(
\ba{c}
\nu_e\\
e
\ea \r)_L,\quad l_R = \l(
\ba{c}
\nu_e\\
e
\ea
\r)_R,\\
\tilde{\phi}&\equiv& \tau_2 \phi^*\tau_2.
\eea
Here $l_L$ is the left-handed lepton which is the $SU(2)_L$ doublet
with the $U(1)$ charge $-1$, while $l_R$ is the right-handed lepton
which is the $SU(2)_R$ doublet with the $U(1)$ charge $-1$.
Three Higgs fields $\phi, \Delta_L$ and $\Delta_R$ carry the 
following $SU(2)_L, SU(2)_R$ and $U(1)_{B-L}$ quantum numbers,
respectively;
\be
\phi(1/2, 1/2^*,0),\quad \Delta_L(1,0,2),\quad \Delta_R(0,1,2).
\ee
The representations of the fields are conveniently given by the $2 \times 2$
matrices as
\bea
\phi&=& \l(
\ba{cc}
\phi_1^0 & \phi_1^+\\
\phi_2^- & \phi_2^0
\ea \r),\\
\Delta_L&=&\l(
\ba{cc}
\delta_L^+/\sqrt{2} & \delta_L^{++} \\
\delta_L^0 & - \delta_L^+/\sqrt{2}
\ea \r),\quad
\Delta_R = \l(
\ba{cc}
\delta_R^+/\sqrt{2} & \delta_R^{++} \\
\delta_R^0 & - \delta_R^+/\sqrt{2}
\ea \r).
\eea
The gauge fields associated with $SU(2)_L, SU(2)_R$ and $U(1)_{B-L}$
are $W_{\mu}^L, W_{\mu}^R$ and $B_{\mu}$, respectively. This Lagrangian
is invariant under the left-right symmetry defined by
\be
l_L \leftrightarrow l_R,\quad \Delta_L \leftrightarrow \Delta_R,
\quad \phi \leftrightarrow \phi^{\dagger}.
\ee

The most general Higgs potential which is invariant under gauge
transformations and the operation (1$\cdot$10) is given in Ref.[9].
Inspite of the left-right symmetry of the Higgs potential, we can
choose a vacuum such that the vacuum expectation values (VEV) of the Higgs
fields break the left-right symmetry. Let us parametrize the VEV
of each Higgs field as
\be
\bra \phi \ket_0 = \l(
\ba{cc}
\kappa_1/\sqrt{2} & 0 \\
0 & \kappa_2/\sqrt{2} 
\ea
\r),\quad \bra \Delta_{L, R} \ket_0 = \l(
\ba{cc}
0 & 0 \\
v_{L, R} & 0
\ea \r),
\ee
where we assume
\be
|v_L| \ll |\kappa_1|, |\kappa_2| \ll |v_R|
\ee
from the phenomenological reason. Then the charged-lepton mass is given
by
\be
m_{l^+} \simeq \f{1}{\sqrt{2}}|f \kappa_2 + \tilde{f} \kappa_1|,
\ee
whereas the neutrino masses are
\bea
m_{\nu_R} &\simeq& \sqrt{2} \l| h v_R\r|,\\
m_{\nu_L} &\simeq& \sqrt{2} \l|h v_L - \f{f \kappa_1+\tilde{f} \kappa_2}
{4 h v_R}\r|.
\eea
The gauge boson masses are
\bea
m_{W_L}^2 &\simeq& \f{1}{4} f_2^2\l(|\kappa_1|^2+|\kappa_2|^2\r)
\sim m_Z^2,\\
m_{W_R}^2 &\simeq& \f{1}{2} g_2^2 |v_R|^2 \sim M_X^2.
\eea
It is found from (1$\cdot$12) that
$m_{\nu_R}, m_{W_R}$ and $m_X$ are too heavy to observe.

\section{LRSM in $M_4 \times Z_2 \times Z_2$ space}

\subsection{The structure of $Z_2 \times Z_2$ group}

Let the discrete points of the first $Z_2$ be $p_1=(1, -1)$ and
those of the second $Z_2$ be $p_2=(1, -1)$. Then four points of
$Z_2 \times Z_2$ are give by (see Fig.1)
{
\setcounter{enumi}{\value{equation}}
\addtocounter{enumi}{1}
\setcounter{equation}{0}
\renewcommand{\theequation}{\thesection$\cdot$\theenumi\alph{equation}}
\be
g_p=(p_1, p_2),\quad p=0, 1, 2, 3
\ee
that is,
\be
g_0=(1, 1), g_1=(-1, 1), g_2=(1, -1), g_3=(-1, -1).
\ee
\setcounter{equation}{\value{enumi}}}
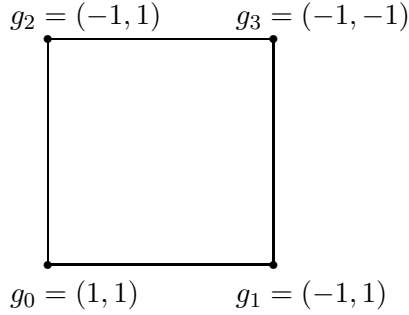
\begin{figure}
\setlength{\unitlength}{1mm}
\begin{center}
\begin{picture}(100,40)(0,0)
\put(35,5){\line(1,0){30}}
\put(35,5){\line(0,1){30}}
\put(65,5){\line(0,1){30}}
\put(35,35){\line(1,0){30}}
\put(30,0){$g_0=(1,1)$}
\put(60,0){$g_1=(-1,1)$}
\put(30,37){$g_2=(-1, 1)$}
\put(60,37){$g_3=(-1,-1)$}
\put(35,5){\circle*{1}}
\put(65,5){\circle*{1}}
\put(35,35){\circle*{1}}
\put(65,35){\circle*{1}}

\end{picture}
\end{center}
\caption{Four discrete points of $Z_2 \times Z_2$.}
\label{fig.1}
\end{figure}
They are subject to the algebraic relations
\be
\ba{c}
g_0 g_p=g_p,\quad g_p g_p=g_0,\quad (p=0, 1, 2, 3)
\vspace{0.3cm}\\
g_1 g_2=g_3,\quad \mbox{and permutations of $(1, 2, 3)$.}
\ea
\ee

We attach the $SU(2) \times U(1)$ internal vector space to each point
of $(x, g_p)$, where $x \in M_4$. The four internal spaces are independent
of each other. Fermionic fields $\psi(x, g_p) \equiv \psi_p$ are assigned
on $(x, g_p) \equiv p$, and they are extended spinors $\times SU(2)$
spinors in each manfold.

\subsection{Fundamental requirements}

We require that our model has the following invariances:
\vx\\
{\bf (i)}
{\it Local invariance under the $SU(2) \times U(1)$ gauge transformation at
each point p.}\\
{\bf (ii)}
{\it The extended $G$-conjugation invariance, which will be explained
in \S 2.5.}\\
{\bf (iii)}
{\it Invariance under the $Z_2 \times Z_2$ discrete transformation.}
\vx

According to (i), the 
$SU(2)$ and $U(1)$ gauge fields in $M_4$ are introduced through
the covariant derivatives of $\psi_p$ at $p$ as
\be
D_{\mu} \psi_p = \l(\p_{\mu}-iA_{p\mu}\r) \psi_p,
\ee
where
\be
A_{p\mu} = B_{p\mu} + W_{p\mu},
\ee
$B_{p\mu}$ being the $U(1)$ gauge field while $W_{p\mu}$ the $SU(2)$ gauge
field.
Connections in $Z_2 \times Z_2$ are
introduced as mapping functions of $\psi_q$ from $q$ to $p$ as
{
\setcounter{enumi}{\value{equation}}
\addtocounter{enumi}{1}
\setcounter{equation}{0}
\renewcommand{\theequation}{\thesection$\cdot$\theenumi\alph{equation}}
\be
\psi_q \to \Phi_{pq} \psi_q
\ee
with the property
\be
\Phi_{pq} = \Phi_{qp}^{\dagger},
\ee
\setcounter{equation}{\value{enumi}}}
$\!\!$where the mapping function
$\Phi_{pq}$ ``parallel-transports'' $\psi_q$ to the fiber on $p$
and will be identified with the Higgs fields.
Here we add the fourth requirement for our model construction:
\vx
\begin{description}
\item[(iv) ]
{\it The mapping functions $\Phi_{pq}$ are given only along each
constituent $Z_2$, namely, along each side of the square depicted in Fig.1:
$0 \leftrightarrow 1,
2 \leftrightarrow 3,
0 \leftrightarrow 2$ and
$1 \leftrightarrow 3$.} (There are no direct connections along the diagonals:
$0 \leftrightarrow 3$ and $1 \leftrightarrow 2$.)
\end{description}
\vx

Under gauge transformations $U_p$ of $\psi_p$, the mapping function
$\Phi_{pq}$ should obey the following rule, 
\be
\Phi_{pq} \to \Phi'_{pq} = U_p \Phi_{pq} U_q^{-1},
\ee
just like in lattice gauge theories.
According to this transformation rule, we can regard
$\Phi_{pq}$ as gauge fields in $Z_2 \times Z_2$.

\subsection{Fermionic Lagrangian}

The fermionic Lagrangian in $M_4 \times Z_2 \times Z_2$ without gauge fields
is assumed to be
\be
{\cal L}_F^0 = i \sum_p \bar{\psi}_p \gamma^{\mu} \p_{\mu} \psi_p
- {\sum_{p,q}}' \kappa_{p+q} \bar{\psi}_p \gamma^{p+q} \psi_q,
\ee
where $\kappa_{p+q}$ are real mass parameters and $\sum'$ implies
the summation over $p$ and $q$ with the constraint $p+q=h=1, 2$ is taken.
The last condition follows since we take into account only of
interactions between nearest-neighbor points in Fig.1.
The Dirac matrices $\gamma^h\ (h=1, 2)$ are
associated with the $Z_2 \times Z_2$ space and the whole set of the
$\gamma$ matrices $\{\gamma^{\mu}, \gamma^h\}$ satisfies the
six-dimensional Clifford algebra. The extra Dirac matrices are chosen as
\be
\gamma^{h} = i \gamma_5 \sigma^h\quad \mbox{($h=p+q=1,2$
and $\gamma_5^{\dagger} = \gamma_5$)},
\ee
where $\sigma^h$ are the Pauli matrices. We have discarded here the term
with $p=q$, which would arise if one naively replaces the
derivative in the continuum part by the difference. This amounts to
proposing a new type of interactin between fermion fields on a
discrete space.\addtocounter{footnote}{-3}\footnote{
The possibility of this type of interaction will be discussed in
a separate paper.} As will be seen below, this choice of interaction is
crucial in our theory.

If we require that the Lagrangian (2$\cdot$7) is locally invariant under
the $SU(2) \times U(1)$ gauge transformations
in $M_4 \times Z_2 \times Z_2$, we should introduce gauge fields
(2$\cdot$4) and $\Phi_{pq} = \Phi_{qp}^{\dagger}$ into ${\cal L}_F^0$ as
\be
{\cal L}_F = i \sum_p \bar{\psi}_p \gamma^{\mu} D_{\mu} \psi_p
- {\sum_{p,q}}' \kappa_h \bar{\psi}_p \gamma^h \Phi_{pq} \psi_q,
\ (h=p+q=1, 2).
\ee
The relevant mapping functions are
\be
\Phi_{01}, \Phi_{23} \ \mbox{for $h=1$,}
\ee
and
\be
\Phi_{20}, \Phi_{31} \ \mbox{for $h=2$.}
\ee
The interaction terms in Eq.(2$\cdot$9)
are composed of $A_{p\mu}$ and $\Phi_{pq}$
as
$$
{\cal L}_I = {\cal L}_1 + {\cal L}_2,
$$
where
\be
\ba{c}
{\cal L}_1 = \sum_p \bar{\psi}_p \gamma \cdot A \psi_p,
\vspace{0.3cm}\\
{\cal L}_2 = -i {\sum_{p,q}}' \kappa_h
\bar{\psi}_p \gamma_5 \sigma^h \Phi_{pq} \psi_q,\ (h=p+q=1,2).
\ea
\ee

\subsection{Superselection rules}

For the fermionic interaction Lagrangian ${\cal L}_I$ we find
two kinds of superselection rules:

(I) Let us define $\Gamma_5$ in the (4+2)-dimensional space by
\be
\Gamma_5=\gamma_5 \sigma_3.
\ee
Then $\Gamma_5=1$ and $\Gamma_5=-1$ components of $\psi_p$ are decoupled
with each other.

(II) Let us define $\nu$ for the group element $g_p=(p_1, p_2)$ by
\be
\nu=p_1 p_2 = \pm 1.
\ee
One can see that the interaction ${\cal L}_1$ does not change sings of
$\nu$ and $\sigma_3$, whereas the interaction ${\cal L}_2$ changes
$\nu \to -\nu$ and $\sigma_3 \to -\sigma_3$. Hence, the product
$\nu \sigma_3$ does not change the sign for both ${\cal L}_1$ and
${\cal L}_2$. Therefore the components of $\psi_p$ with $\nu \sigma_3=1$ and
$\nu \sigma_3=-1$ decouple with each other.

From (I) and (II) we can divide the fermionic field into four sectors.
Let us consider one of them,
\be
\mbox{$\Gamma_5=-1$ and $\nu \sigma_3=1$,}
\ee
for example. Since $\Gamma_5 =\gamma_5 \sigma_3=-1$ for this choice,
the fermionic field $\psi_p$ should be of the form
\be
\psi_p=\l(
\ba{c}
\psi_{pL}
\vspace{0.2cm}\\
\psi_{pR}
\ea \r)
\ba{l}
\ \cdots \sigma_3=+1
\vspace{0.2cm}\\
\ \cdots \sigma_3=-1,
\ea
\ee
where $\psi_L(\psi_R)$ is the left(right)-handed component of $\psi$,
satisfying $\gamma_5 \psi_L=-\psi_L\ (\gamma_5 \psi_R = \psi_R)$.
Then one obtains from $\nu \sigma_3=1$
\bea
\sigma_3=\nu=+1&:&
\psi_0= \l(
\ba{c}
\psi_{0L}\\
0
\ea \r),
\psi_3= \l(
\ba{c}
\psi_{3L}\\
0
\ea \r),\\
\sigma_3=\nu=-1&:&
\psi_1= \l(
\ba{c}
0\\
\psi_{1R}
\ea \r),
\psi_2= \l(
\ba{c}
0\\
\psi_{2R}
\ea \r).
\eea

In terms of these assignments of $\psi_p$, the interacation Lagrangians
can be written as
\bea
{\cal L}_1&=& \bar{\psi}_{0L} \gamma \cdot A_0 \psi_{0L}
+ \bar{\psi}_{1R} \gamma \cdot A_1 \psi_{1R}\nn\\
& & + \bar{\psi}_{2R} \gamma \cdot A_2 \psi_{2R}
+ \bar{\psi}_{3L} \gamma \cdot A_3 \psi_{3L},\\
{\cal L}_2 &=& 
-i \kappa_1\l(\bar{\psi}_0 \gamma_5 \sigma^1 \Phi_{01} \psi_1 
+\bar{\psi}_2 \gamma_5 \sigma^1 \Phi_{23} \psi_3\r)\nn\\
& & -i \kappa_2\l(\bar{\psi}_2 \gamma_5 \sigma^2 \Phi_{20} \psi_0
+\bar{\psi}_3 \gamma_5 \sigma^2 \Phi_{31} \psi_1\r) + {\rm h.c.}\nn\\
&=&
-i \kappa_1\l(\bar{\psi}_{0L} \gamma_5 \Phi_{01} \psi_{1R} 
+\bar{\psi}_{2R} \gamma_5 \Phi_{23} \psi_{3L}\r)\nn\\
& & + \kappa_2\l(\bar{\psi}_{2R} \gamma_5 \Phi_{20} \psi_{0L}
-\bar{\psi}_{3L} \gamma_5 \Phi_{31} \psi_{1R}\r) + {\rm h.c.}
\eea

\subsection{Extended $G$-conjugation}

Let us return to the section in which we have not selected one of the sectors.
The extended $G$-conjugation of $\psi_p$ is defined by
\be
\psi_p \to U_G \psi_p U_G^{-1} = \lambda_p (-i \sigma_2)(-i \tau_2) \psi_p^c
\equiv \lambda_p \psi_p^G,
\ee
where $\psi_p^G$ is the charge conjugate field of $\psi_p$ and $\lambda_p$ is
the $G$-parity of $\psi_p$. We set $\lambda_p = 1$ in
the following, for simplicity, and hence
\be
\psi_p^{GG}= \psi_p.
\ee
For each of $\sigma_3$-components of $\psi_p$, we have
\be
\l(
\ba{c}
\psi_{p+}\\
\psi_{p-}
\ea \r)^G = 
\l(
\ba{c}
-\psi_{p-}^g\\
\psi_{p+}^g
\ea \r),
\ee
where $\psi_{p\pm}^g$ is the conventional (i.e., without $-i \sigma_2$)
$G$-conjugation of $\psi_{p\pm}$ and
\be
\psi_{p\pm}^{gg} = - \psi_{p\pm}.
\ee
Note the following formulae
\bea
\bar{\psi}_p(1, \gamma_5)\psi_q 
&=& 
\overline{\psi_p^G}(1, \gamma_5)\psi_q^G,\\
\bar{\psi}_p(\gamma_{\mu}, \vec{\tau}, \sigma_h)\psi_q 
&=& 
- \overline{\psi_q^G}(\gamma_{\mu}, \vec{\tau}, \sigma_h)\psi_p^G
\eea
and
\be
\Omega \to U_G \Omega U_G^{-1}=\lambda_{\Omega} (-i \tau_2) \Omega^*
(-i \tau_2)^{-1} = \lambda_2 \tilde{\Omega} \equiv \lambda_{\Omega} \Omega^G,
\ee
where $\Omega$ stands for the gauge field $A_{p \mu}$ or $\Phi_{pq}$.
The $G$-parity of each gauge field should be $\lambda_{\Omega} = -1$,
according to Eqs. (2$\cdot$24) $\sim$ (2$\cdot$26),
for the interaction Lagrangian ${\cal L}_I$ be $G$-invariant.

\subsection{Kinetic terms of Higgs fields}

Let us calculate the curvature associated with paths in Fig.2.
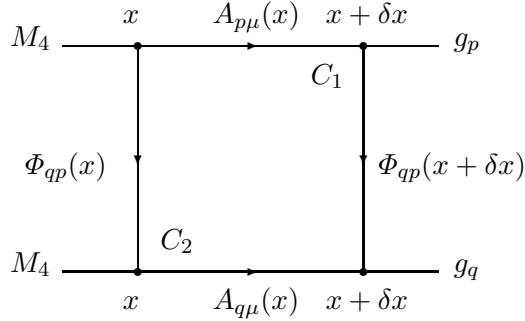
\begin{figure}
\setlength{\unitlength}{1mm}
\begin{center}
\begin{picture}(100,40)(0,0)
\put(25,5){\line(1,0){50}}
\put(25,35){\line(1,0){50}}
\put(35,5){\line(0,1){30}}
\put(65,5){\line(0,1){30}}
\put(35,30){\vector(0,-1){11}}
\put(65,30){\vector(0,-1){11}}
\put(40,5){\vector(1,0){11}}
\put(40,35){\vector(1,0){11}}
\put(18,5){$M_4$}
\put(18,35){$M_4$}
\put(77,5){$g_q$}
\put(77,35){$g_p$}
\put(20,18){$\Phi_{qp}(x)$}
\put(67,18){$\Phi_{qp}(x+\delta x)$}
\put(45,38){$A_{p\mu}(x)$}
\put(45,0){$A_{q\mu}(x)$}
\put(58,30){$C_1$}
\put(38,8){$C_2$}
\put(33,0){$x$}
\put(33,38){$x$}
\put(60,0){$x+\delta x$}
\put(60,38){$x+\delta x$}
\put(35,5){\circle*{1}}
\put(35,35){\circle*{1}}
\put(65,5){\circle*{1}}
\put(65,35){\circle*{1}}

\end{picture}
\end{center}
\caption{The fermion field $\psi_p(x)$ is mapped from $(x, g_p)$ to
$(x+\delta x,g_q)$ through two paths $C_1$ and $C_2$. The
difference between two images defines the curvature $D_{\mu} \Phi_{qp}$.}
\label{fig.2}
\end{figure}
The fermionic field $\psi_p(x)$ is mapped from $(x, g_p)$ to
$(x+\delta x, g_q)$ through paths $C_1$ and $C_2$. The two images are
\bea
\psi(C_1) &=& \Phi_{qp}(x+\delta x)\l[1+ i A_{p\mu}(x) \delta x^{\mu}\r]
\psi_p(x),\\
\psi(C_2) &=&\l[1+ i A_{q\mu}(x) \delta x^{\mu}\r]\Phi_{qp}(x)
\psi_p(x).
\eea
The difference between them yields
\be
\psi(C_1) - \psi(C_2) = D_{\mu} \Phi_{qp}(x) \delta x^{\mu} \psi_p(x),
\ee
where
\be
D_{\mu} \Phi_{qp}= \p_{\mu}\Phi_{qp}-i
\l[A_{q\mu}\Phi_{qp}-\Phi_{qp} A_{p\mu}\r].
\ee
The covariant derivative of $\Phi_{qp}$ is, therefore, just the curvature
associated with Fig.2.

\subsection{Identification of fields}

We first note that our system is $G$-invariant and also has a
symmetry under the translation $p \to p+h\ (h=1, 2)$. The latter translation
symmetry is obvious for $i\sum_p \bar{\psi}_p \gamma^{\mu} D_{\mu} \psi_p$
and $\sum_{p,q}|D_{\mu} \Phi_{qp}|^2$. However, it is not so clear that
${\cal L}_2$ has this invariance. Let us write ${\cal L}_2$ explicitly
again to prove this;
\bea
\sum_{p,q}\,' \kappa_h \bar{\psi}_p \gamma^h \Phi_{qp} \psi_p
&=& \kappa_1\l(\bar{\psi}_0 \gamma^1 \Phi_{01} \psi_1
+ \bar{\psi}_2 \gamma^1 \Phi_{23} \psi_3\r)\nn\\
& & + \kappa_2\l(\bar{\psi}_2 \gamma^2 \Phi_{20} \psi_0
+ \bar{\psi}_3 \gamma^2 \Phi_{31} \psi_1 \r) + \mbox{h.c.}
\eea
Equation (2$\cdot$32) is invariant
under the interchange $0 \leftrightarrow 1,\ 2 \leftrightarrow 3$,
with suffices of $\kappa$ and $\gamma$ kept fixed, due to the
Hermiticity $\Phi_{pq} = \Phi_{qp}^{\dagger}$. This is the translational
symmetry of the type $p \to p+1$. It is also invariant under
$0 \leftrightarrow 2,\ 1 \leftrightarrow 3$, namely, the
translational symmetry of the type $p \to p+2$.

Consequently our system is invariant under the $G$-conjugation
combined with the translation $p \to p+h\ (h=1, 2)$, i.e.,
\be
\psi_p \to \psi_{p+h}^G,\quad A_p \to -A_{p+h}^G,\quad
\Phi_{pq} \to -\Phi_{p+h, q+h}^G.
\ee
From Eqs.(2$\cdot$16 $\sim$ 18) and (2$\cdot$22), the first one $\psi_p \to
\psi_{p+h}^G$ is reduced to
\be
\psi_{pL} \to -\psi_{p+h, R}^g\ (p=0,3),\quad
\psi_{pR} \to \psi_{p+h, L}^g\ (p=1,2).
\ee

Now we have too many independent fields to reconstruct the LRSM
from our geometric formulation of gauge theory in $M_4 \times Z_2 \times
Z_2$. Some of the fields must be removed. In view of the invariance
(2$\cdot$33) and (2$\cdot$34), it is natural to assume
\be
\psi_p = \psi_{p+2}^G,\quad A_p = - A_{p+2}^G, \quad
\Phi_{pq} = - \Phi_{p+2,q+2}^G.
\ee
Another possible choice $\psi_p=\psi_{p+1}^G$, etc. will lead to the same
LRSM. Let the independent fields be $\psi_{0L}, \psi_{1R}, A_0, A_1,
\Phi_{01}, \Phi_{20}, \Phi_{31}$. Then we set them in familiar notations
introduced in \S 1,
\bea
\psi_{0L}&=&l_L = \l(
\ba{c}
\nu_L\\
e_L
\ea \r),\quad \psi_{1R}=l_R=\l(
\ba{c}
\nu_R\\
e_R
\ea \r),\nn\\
A_0&=&B_0 + W_L,\quad A_1=B_1+W_R,\nn\\
\Phi_{01}&=& \phi,\nn\\
\Phi_{20} &=& \Delta_L,\quad \Phi_{31}=\Delta_R,
\eea
(see Fig.3).
%
%
\begin{figure}
\setlength{\unitlength}{1mm}
\begin{center}
\begin{picture}(100,40)(0,0)
\put(35,5){\line(1,0){30}}
\put(35,5){\line(0,1){30}}
\put(65,5){\line(0,1){30}}
\put(35,35){\line(1,0){30}}
\put(37,7){$g_0$}
\put(60,7){$g_1$}
\put(37,30){$g_2$}
\put(60,30){$g_3$}
\put(35,5){\circle*{1}}
\put(65,5){\circle*{1}}
\put(35,35){\circle*{1}}
\put(65,35){\circle*{1}}
\put(35,11){\vector(0,1){11}}
\put(65,11){\vector(0,1){11}}
\put(60,5){\vector(-1,0){11}}
\put(60,35){\vector(-1,0){11}}
\put(49,0){$\phi$}
\put(47,38){$-\tilde{\phi}$}
\put(27,20){$\Delta_L$}
\put(67,20){$\Delta_R$}
\put(30,0){$l_L$}
\put(30,37){$l_L^g$}
\put(70,0){$l_R$}
\put(70,37){$-l_R^g$}
\end{picture}
\end{center}
\caption{Assignments of fields on $Z_2 \times Z_2$.}
\label{fig.3}
\end{figure}
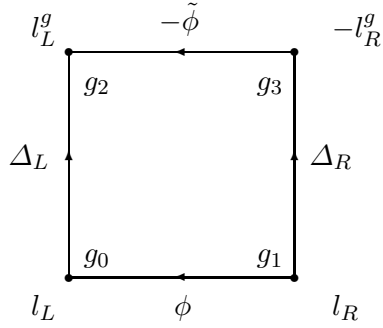
From Eq.(2$\cdot$35) other fields are given by
\bea
\psi_{2R}&=&\psi_{0L}^g = l_L^g = -i \tau_2 C^{-1}\bar{l}_L^T,\nn\\
\psi_{3L}&=&-\psi_{1R}^g = -l_R^g = i \tau_2 C^{-1}\bar{l}_R^T,\nn\\
A_2&=& -A_0^G = - \tilde{A}=-(\tilde{B}_0+\tilde{W}_L)=-B_0+W_L,\nn\\
A_3&=& -A_1^G =-\tilde{A}_1=-B_1+W_R,\nn\\
\Phi_{23}&=& -\Phi_{01}^G=-\phi^G=-\tilde{\phi}.
\eea
Here we have still two kinds of $U(1)$ gauge fields $B_0$ and $B_1$, which
are defined independently on $M_4 \times \{g_0\}$ and $M_4 \times \{g_1\}$.
If we identify $B_0$ with $B_1$, i.e., $B_0=B_1=B$, we have the standard
LRSM, except for the Higgs potential. This will be studied in the
next subsection.

In order to normalize the kinetic term, we rescale the fermions as
$\psi_p \to \f{1}{\sqrt{2}} \psi_p$. The $\gamma_5$ factor in
${\cal L}_2$ can be removed by redefining $\psi_p$ as
\be
e^{i\pi \gamma_5/4}\psi_p \to \psi_p,
\ee
where the use has been made of the identity $i\gamma_5 
= e^{i \pi \gamma_5/2}$. The interaction Lagrangians ${\cal L}_1$ and
${\cal L}_2$ in Eqs.(2$\cdot$19) and (2$\cdot$20) are reduced to
\bea
{\cal L}_1&=& \f{1}{2} \bar{l}_L \gamma \cdot A_0 l_L + 
\f{1}{2} \bar{l}_R \gamma \cdot A_1 l_R + 
\f{1}{2} \overline{l_L^g} \gamma \cdot A_2 l_L^g + 
\f{1}{2} \overline{l_R^g} \gamma \cdot A_3 l_R^g\nn\\
&=& \bar{l}_L \gamma \cdot (B+W_L) l_L 
+ \bar{l}_R \gamma \cdot (B+W_R) l_R,\\
{\cal L}_2 &=& - \kappa_1 \l(\bar{l}_L \phi l_R + \bar{l}_R \phi l_L\r)\nn\\
& &- \f{1}{2} i\kappa_2\l(
\overline{l_L^g} \Delta_L l_L + \overline{l_R^g} \Delta_R l_R
- \bar{l}_L \Delta_L^{\dagger} l_L^g - \bar{l}_R \Delta_R^{\dagger} l_R^g\r).
\eea
Covariant derivatives of Higgs fields (2$\cdot$30) are
\bea
D \phi&=& \p \phi -i(W_L \phi-\phi W_R),\\
D \Delta_L&=& \p \Delta_L +2iB \Delta_L-i(W_L \Delta_L-\Delta_L W_L),\\
D \Delta_R&=& \p \Delta_R +2iB \Delta_R-i(W_R \Delta_R-\Delta_R W_R)
\eea

\subsection{Higgs potential}

Let us calculate the curvature associated with the paths in Fig.4,
where mapping functions $\Phi_{pq} = \Phi_{pq}^{\dagger}$ are still
independent of each other.
%
%
\begin{figure}
\setlength{\unitlength}{1mm}
\begin{center}
\begin{picture}(100,40)(0,0)
\put(35,5){\line(1,0){30}}
\put(35,5){\line(0,1){30}}
\put(65,5){\line(0,1){30}}
\put(35,35){\line(1,0){30}}
\put(30,1){$g_0$}
\put(67,1){$g_1$}
\put(30,37){$g_2$}
\put(67,37){$g_3$}
\put(35,12){\vector(0,1){10}}
\put(65,12){\vector(0,1){10}}
\put(58,5){\vector(-1,0){10}}
\put(58,35){\vector(-1,0){10}}
\put(47,1){$\Phi_{01}$}
\put(47,38){$\Phi_{23}$}
\put(25,20){$\Phi_{20}$}
\put(67,20){$\Phi_{31}$}
\put(35,5){\circle*{1}}
\put(65,5){\circle*{1}}
\put(35,35){\circle*{1}}
\put(65,35){\circle*{1}}
\end{picture}
\end{center}
\caption{The mapping functions between points of $Z_2 \times Z_2$. The
holonomy associated with two paths defines the curvature $G_{p+3,p}$.}
\label{fig.4}
\end{figure}
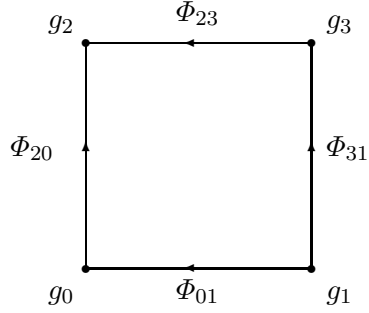
Identification of the fields will be introduced later. The fermion
field $\psi_p$ on $(x, g_p)$ is mapped from $(x, g_p)$ to $(x, g_{p+3})$
through two paths $p \to p+1 \to p+3$ and $p \to p+2 \to p+3$.
The difference between the two mappings, namely the holonomy associated
with the paths, yields the curvature
\be
G_{p+3, p} \equiv \Phi_{p+3, p+1} \Phi_{p+1, p}-
\Phi_{p+3, p+2} \Phi_{p+2, p} = G_{p, p+3}^{\dagger}.
\ee

There is also a different type of curvature arising from Fig. 5.
%
%
\begin{figure}
\setlength{\unitlength}{1mm}
\begin{center}
\begin{picture}(100,20)(0,0)
\put(35,8){\line(1,0){30}}
\put(35,12){\line(1,0){30}}
\put(65,10){\oval(4,4)[r]}
\put(46,8){\vector(1,0){5}}
\put(55,12){\vector(-1,0){5}}
\put(35,10){\circle*{1}}
\put(67,10){\circle*{1}}
\put(30,9){$g_p$}
\put(68,9){$g_{p+h}$}
\put(50,1){$C_3$}
\put(50,14){$C_4$}
\end{picture}
\end{center}
\caption{The path defining the curvature $F_{p,p+h,p}$.}
\label{fig.5}
\end{figure}
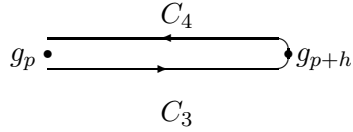
This is the same type as in the Weinberg-Salam electroweak model.
That is, $\psi_p$ is compared with its image $\psi_p(C_3 \cdot C_4)$, which is
obtained by mapping $\psi_p$ through the path $C_3 \cdot C_4$, i.e.,
\be
\psi_p(C_3 \cdot C_4) - \psi_p =\l(\Phi_{p,p+h} \Phi_{p+h, p}-1\r) \psi_p.
\ee
This defines the curvature
\be
F_{p, p+h, p} \equiv \Phi_{p,p+h} \Phi_{p+h, p}-1,\quad (h=1, 2).
\ee
These curvatures take finite values in a discrete space although they
vanish in a continuous space.

Since Higgs fields are gauge fields, the Higgs potential must be of the
Yang-Mills type and gauge-invariant.
Gauge invariant combinations of curvatures are as follows:
\bea
A_{p:h,k}&\equiv& \f{1}{2} \tr \l(F_{p,p+h,p} F_{p,p+k,p}\r),
\ (p=0,1,2,3;\ h,k=1,2)\\
B_{p,q:h,k}&\equiv& \f{1}{2} \tr \l(F_{p,p+h,p}\r)
\f{1}{2} \tr\l( F_{q,q+k,q}\r),
\ (p,q=0,1,2,3;\ h,k=1,2)\\
\Gamma_p&\equiv& \f{1}{2} \tr \l(G_{p,p+3} G_{p+3, p}\r).\ (p=0, 1)
\eea
Then the Higgs potential is written as a linear combinations of them,
\bea
V&=&
\alpha_{11} \sum_{p=0,2} A_{p:1,1} +
\alpha_{12} \sum_{p=0,1} A_{p:2,2} +
\alpha_{2} \sum_{p=0}^3 A_{p:1,2} \nn\\
& &+
\beta_{11} \sum_{p=0,2} B_{p,p:1,1} +
\beta_{12} \sum_{p=0,1} B_{p,p:2,2} +
\beta_{2} \sum_{p=0}^3 B_{p,p:1,2}\nn\\
& & +
\beta_{31} B_{0,3:1,1}+\beta_{32} B_{0,3:2,2}\nn\\
& & + \gamma\ (\Gamma_0+\Gamma_1),
\eea
where $\alpha, \beta$ and $\gamma$ are arbitrary real parameters.

In the following, we use following formulae for the $G$-conjugation
$\tilde{A} = \tau_2 A^* \tau_2$:
\bea
\tilde{\tilde{A}} &=& A,\nn\\
\tilde{\tau}_i &=& - \tau_i,\ \mbox{for $i=1, 2, 3$},\nn\\
\widetilde{AB} &=& \tilde{A} \tilde{B},\nn\\
\tr(\tilde{A}) &=&\tr (A^{\dagger}).
\eea
After identifications of fields (2$\cdot$35) and (2$\cdot$36),
the Higgs fields
$\Delta_{L,R}$ have a form $\Delta = \sum_{i=1}^3 \tau_i \Delta_i$, and
hence
\be
\tilde{\Delta}_{L, R} = - \Delta_{L, R}^{\dagger}.
\ee
From these formulas one obtains the identity
\be
\Gamma_0 = \Gamma_1,
\ee
since
\begin{eqnarray*}
2\Gamma_1&=& \tr \l(G_{12} G_{21}\r)\\
&=& \tr
\l\{ \l(\Phi_{13} \Phi_{32}-\Phi_{10} \Phi_{02}\r)
\l(\Phi_{20} \Phi_{01}-\Phi_{23} \Phi_{31}\r)\r\}\\
&=& \tr
\l\{ \l(-\Delta_R^{\dagger}\tilde{\phi}^{\dagger}-
\phi^{\dagger} \Delta_L^{\dagger}\r)
\l(\Delta_L \phi + \tilde{\phi} \Delta_R\r)\r\}\\
&=& \tr
\l\{ \l(\tilde{\Delta}_R \tilde{\phi}^{\dagger}+
\phi^{\dagger} \tilde{\Delta}_L^{\dagger}\r)
\l(-\tilde{\Delta}_L^{\dagger} \phi - \tilde{\phi} \tilde{\Delta}_R^{\dagger}
\r)\r\}\\
&=& \tr
\l\{ \l(\Delta_R \phi^{\dagger}+
\tilde{\phi}^{\dagger} \Delta_L\r)^{\sim}
\l(-\Delta_L^{\dagger} \tilde{\phi} - \phi \Delta_R^{\dagger}\r)^{\sim}\r\}\\
&=& \tr
\l\{\l(-\tilde{\phi}^{\dagger} \Delta_L - \Delta_R \phi^{\dagger}\r)
\l(\phi \Delta_R^{\dagger} + \Delta_L^{\dagger} \tilde{\phi}\r)\r\}
\end{eqnarray*}
and
\begin{eqnarray*}
2\Gamma_0&=&\tr\l(G_{03}G_{30}\r)\\
&=& \tr\l\{\l(\Phi_{02} \Phi_{23}-\Phi_{01} \Phi_{13}\r)
\l(\Phi_{31} \Phi_{10}-\Phi_{32}\Phi_{20}\r)\r\}\\
&=& \tr\l\{
\l(-\Delta_L^{\dagger} \tilde{\phi}-\phi \Delta_R^{\dagger}\r)
\l(\Delta_R \phi^{\dagger} + \tilde{\phi}^{\dagger} \Delta_L\r)\r\}\\
&=& 2\Gamma_1.
\end{eqnarray*}
Similarly it can be shown that
\bea
\tr\l(\tilde{\phi}\tilde{\phi}^{\dagger} \Delta_L \Delta_L^{\dagger}\r)
&=& \tr\l(\phi \phi^{\dagger} \tilde{\Delta}_L \tilde{\Delta}_L^{\dagger}\r)
^{\sim}
= \tr\l(\phi \phi^{\dagger} \Delta_L^{\dagger} \Delta_L\r)^{\dagger}
= \tr\l(\phi \phi^{\dagger} \Delta_L^{\dagger} \Delta_L\r),\nn\\
\\
\tr\l(\tilde{\phi}^{\dagger}\tilde{\phi}
\Delta_R \Delta_R^{\dagger}\r)&=&
\tr\l(\phi^{\dagger} \phi \Delta_R^{\dagger} \Delta_R \r),
\eea
from which one can prove the equality
\be
\tr \l(F_{010} F_{020} \r) = 
\tr \l(F_{232} F_{202} \r) = 
\tr\l\{\l(\f{1}{2} \phi \phi^{\dagger}
-1\r)\l(\Delta_L^{\dagger} \Delta_L-1\r)\r\}.
\ee
It can be also shown that
\be
\tr \l(F_{101} F_{131} \r) =
\tr \l(F_{323} F_{313} \r) = \tr\l\{ \l( \f{1}{2} \phi^{\dagger} \phi
-1 \r)\l( \Delta_R^{\dagger} \Delta_R -1\r)\r\}.
\ee

Taking account of these identities the Higgs potential turns out to be
of the form
\bea
V&=& \alpha_{11} \tr \l(F_{\phi}^2 \r)+
\f{1}{2} \alpha_{12} \l[\tr\l(F_L^2 \r)+\tr \l(F_R^2 \r)\r]
+ \alpha_{2} \l[\tr\l(F_{\phi} F_L \r)+\tr \l(F'_{\phi}F_R \r)\r]\nn\\
& &+\f{1}{2} \l(\beta_{11}+\f{1}{2}\beta_{31} \r)
\l(\tr F_{\phi}\r)^2
+ \f{1}{4} \beta_{12} \l[ \l(\tr F_L\r)^2 + \l(\tr F_R\r)^2\r]\nn\\
& &+\f{1}{2} \beta_2 \l(\tr F_{\phi}\r)
\l[\l(\tr F_L\r)+\l(\tr F_R\r)\r] + \f{1}{4} \beta_{32} \l(\tr F_L\r)
\l(\tr F_R \r)\nn\\
& & +\gamma \tr \l(G_{03} G_{30}\r),
\eea
where
\bea
F_{\phi}&=& F_{010} = \f{1}{2} \phi \phi^{\dagger}-1,\quad
F'_{\phi} = F_{101} = \f{1}{2} \phi^{\dagger} \phi - 1,\nn\\
F_{L}&=& F_{020} = \Delta_L^{\dagger}\Delta -1,\quad
F_R = F_{131} = \Delta_R^{\dagger} \Delta_R - 1,\nn\\
G_{30}&=& G_{03}^{\dagger} = \Delta_R \phi^{\dagger} + 
\tilde{\phi}^{\dagger} \Delta_L.
\eea
The factor $1/2$ in $F_{\phi}$ is necessary when we identify fields
$\Phi_{01}$ and $\Phi_{23}$ with $\phi$ and $-\tilde{\phi}$, respectively.
In this case, two Higgs kinetic terms, $\tr |D \Phi_{01}|^2$ and 
$\tr |D \Phi_{23}|^2$, agree with each other, so that we should
redefine the field $\phi$ as $\phi \to \phi/\sqrt{2}$. As for
other fields $W_L, W_R$ and $B$ the same thing happens, but one can
rescale their coupling constans by the factor $1/\sqrt{2}$.

The Higgs potential thus obtained is left-right symmetric under the
operation (1$\cdot$10). This property comes from our fundamental
requirement (iii) in \S 2.2. Since our Higgs fields $\phi, \Delta_L$
and $\Delta_R$ are dimensionless, they should be redefined so as to
be $[\mbox{Higgs fields}] = L^{-1}$. In the redefinition of the fields we need
three new parameters. Consequently our Higgs potential (2$\cdot$58)
contains eleven free parameters.

If the symmetry between 1 and 2, namely the exchange of two $Z_2$'s, 
is required, one obtains
\bea
\alpha_{11}&=&\alpha_{12} \equiv \alpha_1,\nn\\
\beta_{11}& =& \beta_{12} \equiv \beta_1,\nn\\
\beta_{31} &=& \beta_{32} \equiv \beta_3.
\eea

\section{Conclusion}

The LRSM with gauge group $SU(2)_L \times SU(2)_R \times U(1)_{B-L}$ has 
been reconstructed from the geometric point of view of gauge theory
in $M_4 \times Z_2 \times Z_2$. We have started with the fundamental
requirements (i) $\sim$ (iv) stated in \S 2.2. Our new results are then
summarized as follows:
\begin{enumerate}
\item Three Higgs fields $\phi, \Delta_L$ and $\Delta_R$
are gauge fields in $Z_2 \times Z_2$.
\item The Higgs potential, therefore, should be of Yang-Mills
type, i.e., it is given by (2$\cdot$58) as a sum of 
$$
\tr |\mbox{curvatures of Higgs fields}|^2,
$$
which contains eleven free parameters. This should
be compared with the general Higgs potential given by Deshpande
{\it et al.},\cite{ref:9} which contains eighteen parameters.
\item As a result we have obtained the property that our Higgs
potential is left-right-symmetric under the operation (1$\cdot$10).
This property comes from our fundamental requirement (iii) in \S 2.2,
that is, our model should be invariant under the discrete
$Z_2 \times Z_2$ transformation.
\item The Yukawa coupling Lagrangian ${\cal L}_2$ given by
(2$\cdot$40) does not contain the term $\bar{l}_L \tilde{\phi} l_R
+ {\rm h.c.}$, which appears in ${\cal L}_Y$ of (1$\cdot$3) in the standard
LRSM.
\item In \S 2.7, we have identified the $U(1)$ gauge field $B_0$
with the other one $B_1$, i.e., $B_0=B_1=B$. However, if we do not
identify them, we have another type of LRSM with two kinds of $U(1)$
gauge fields, $B_0=B_L$ and $B_1 = B_R$.
\end{enumerate}
\vx

Actual phenomenological analyses of the latter LRSM with two $U(1)$
gauge fields and of obtaining
the left-right asymmetric vacuum based on our Higgs potential will
be left for future study.

In conclusion, our approach led to the LRSM quite successfully
and the geometrical structure of this model has been better understood
compared to other works based on NCG. Finally we would like to
thank M.~Kubo for discussions.

\end{document}